# Empirical investigation of key business factors for digital game performance

Saiqa Aleem (Western University, Canada) Luiz Fernando Capretz (Western University, Canada) Faheem Ahmed (Thompson River University, Canada)

ARTICLE INFO

ABSTRACT

Game development is an interdisciplinary concept that embraces software engineering, business, management, and artistic disciplines. This research facilitates a better understanding of the business dimension of digital games. The main objective of this research is to investigate empirically the effect of business factors on the performance of digital games in the market and to answer the research questions asked in this study. Game development organizations are facing high pressure and competition in the digital game industry. Business has become a crucial dimension, especially for game development organizations. The main contribution of this paper is to investigate empirically the influence of key business factors on the business performance of games. This is the first study in the domain of game development that demonstrates the interrelationship between key business factors and game performance in the market. The results of the study provide evidence that game development organizations must deal with multiple business key factors to remain competitive and handle the high pressure in the digital game industry. Furthermore, the results of the study support the theoretical assertion that key business factors play an important role in game business performance.

## 1. Introduction

Over the past three decades, digital games have emerged as an important part of media and global entertainment. The digital game sector is creative, dynamic, pervasive, and exciting. The social media revolution and ever-increasing Internet expansion are driving phenomenal growth for the digital game segment in particular and are creating a huge multimedia business worth billions of dollars. The digital game sector, especially video games, is expected to grow by up to USD 112 billion in sales by 2015 as reported by Gartner Research [1], and overall growth of the digital game sector is expected to reach USD 82.4 billion by 2015 [2]. Digital game development organizations are looking at new ways to improve existing user experiences, to engage a broad range of consumers, to update their business models, and to include emerging technologies in their development processes. The digital game sector has been identified as a significant contributor to economic growth by many countries around the world, and these countries have embraced aggressive action plans for its growth [3, 4]. In the digital game industry, Kerr [5] identified four distinct segments: standard PC, console, casual, and massively multiplayer online games. Organizations in each segment have a different culture for production and entry to market and are structured differently. Game development organizations are directly or indirectly involved in various activities from a game's inception to its consumption. These main activities in general terms, regardless of game genre or particular segment, can be grouped together as: a) development or production; b) commercialization or publishing; c) distribution; and d) customer engagement. Production or development is a major multidisciplinary activity that involves merging of technical and creative disciplines. The development process involves planning, design, development, and test phases and is a kind of iterative process. The publishing activity involves either developing games in-house and outsourcing publishing of developed games or else purchasing of games from independent studios by publisher companies. Outsourced publishing activities can include data hosting, billing, marketing and advertising, intellectual property management, and analytics. Distribution activity is different for online and packaged games. For online games, intermediaries like virtual windows can be used for their distribution. For packaged games, distribution includes handling and packaging of games as well as marketing and logistics. Distribution activities can also be outsourced by game companies. Customer engagement includes customer support activity; some companies also outsource customer support to achieve cost savings.

However, the main development activities in the digital game sector especially from business perspective include elements of the game development value chain, such as technical and creative development, manufacturing of hardware/ console platforms, and game publishing. Distribution can be carried out in a number of ways, including mobile, traditional retail, online, cloud, and download, after distribution, it also involves customer engagement and community management activities. Game development organizations have also outsourced some of their support services activities such as data hosting, information security, marketing and advertising, billing services, and piracy protection. The number of dimensions involved, such as types of end-user devices or platforms, game genres, channels for publication, and emerging revenue models in the digital game sector, make this sector highly fragmented. It is important for any type of business domain to identify its key important factors that help them to excel in that particular field. The key business factors vary from domain to domain depending upon their business operation. Digital game is kind of software product and it is intangible in nature. According to Levitt [6], intangible products are highly people-intensive in their delivery methods and production and business management become more critical for them as compared to tangible products. Moreover, digital game industry growth is tremendous and it became crucial to identify key important business factors that help organization in digital game industry to reach their maximum potential.Game development organizations must target all these dimensions to retain and maximize their consumers. The digital game industry has shown economic potential in both the entertainment and software industries [2].

## 1.1 Research Motivation

Organizations involved in the game development business are facing stiff competition and high consumer expectations because more and more development companies are entering the digital game industry day by day. The main research motivation behind this study is the rapid and continual changes in technology and the severity of competition in game development organization and it not only affect the business, but also have a great impact on development activities. Actually, the game industry has high economic potential and generates million-dollar projects, it sets high limits and standards for game performance as well as putting high pressure on organizations. To deal with this severe competition and high pressure, game development organizations must make important decisions quickly regarding different business activities because this has become important for financial growth and business performance. Organizations in the digital game industry must respond quickly to changes in the business and technology environment, and if they fail to respond appropriately, then they will not survive long. There are many examples of commercial failure in digital game industry and the most popular one is known as video game crash of 1983 [7]. According to Burnham [7], an expensive low quality games with poor business strategies were flooded in North America. They resulted in complete destruction of US digital game market. Also, Sellers [8] stated that the extra-terrestrial (E.T.) video game and Pac Man for Atrai 2600 were two examples that contributed to the failure. Most of the failures in the digital game market such as Commodore 64 Games System, Nitendo 64 DD, Philips Cdi, Shenmue, Sonic Boom: Rise of lyric etc. [9], were due to poor business strategies including market orientation, consumer satisfaction, monetization strategy, time to market etc.

Especially in game development organizations, business becomes the most important factor due to severe competition, the fragmented nature of the business, and the poor software engineering practices used by most companies [5]. Identification of key factors to handle high pressure and achieve targeted business and game performance has become highly important. However, no studies that address the important factors in digital game business performance have been published in the literature. The main contribution of this empirical study is to investigate comprehensively the interrelationship among key business factors and game performance in the market. This study also provides an understanding of the influence of the key factors identified by showing empirically how they impact the business organization and digital game performance.

The rest of the paper is organized as follows: Section 2 provides the research methodology, and Section 3 describes the results and analysis. Section 4 presents a discussion, and Section 5 concludes the study.

## 2. Literature Review

Key business performance factors for digital game organizations are the least addressed area in game development research. The business model for each segment of the game industry is different, and each segment has a different percentage of the revenue share [10]. From a review of the literature, various factors have been identified that contribute to game business performance. The identified factors and the related literature review are described in the following sub-sections.

### 2.1  Customer Satisfaction/Loyalty

The digital game industry (DGI) is facing dramatic changes because it views customer satisfaction as winning over players for their games. The classical definition of customer satisfaction given by Oliver [11] is "pleasurable fulfillment response toward a good, service, benefit, or reward". Customer satisfaction must be an integral part of the organization and is a financial metric that can be used to measure business performance. However, the relationship between business performance and customer satisfaction is not always clear. Zeithaml [12] highlighted three problems in measuring this relationship: a) the time lag between measuring improvement in profit and customer satisfaction; b) other variables that influence an organization's profits, such as marketing, price, and competition; and c) other variables such as organizational behavioral issues that should be included when measuring the relationship. A positive relationship between customer satisfaction and organizational performance has been reported by many researchers in different industries [13, 14, 15, 16, 17], but few have explored this relationship in the DGI. Some researchers [18, 19, and 20] have also highlighted that higher customer satisfaction in any organization is strongly correlated with higher market growth, proving the strong relationship between customer loyalty and customer retention. The DGI has given a lower priority to customer service for its product (the game) and tends to treat it as a commodity. Often, when players do not obtain a response to their problems, they become disappointed. Johnson [21] explored the aspect of customer service in the DGI. He used the critical incidents technique to examine customer services incidents in the game industry and identified negative and positive customer service experiences. The results of this study provided directions for management that helped them with resource allocation, especially in those areas that provided maximum customer satisfaction and dissatisfaction. Based on this analysis, management could take proper measures to ensure maximum customer satisfaction. In commercial games, the concept of customer satisfaction has a very important place. Lu and Wang [22] explored the factors of online game addiction and the role of addiction in online gamer loyalty and customer satisfaction. The results indicated that addiction plays an important role in customer loyalty and satisfaction. Wiele *et al*. [23] investigated the relationship of customer satisfaction and business performance data within an organization. The results showed empirical evidence that there is a positive relationship between customer satisfaction and business performance. In the literature, only a few researchers have explored the customer satisfaction aspect of the DGI.

### 2.2  Market Orientation

Market orientation plays a significantly important role in extensively market-driven DGI. Market orientation involves the study of customers and competitors in the market and deals with the interpretation, acquisition, or use of information about them. The concept of market orientation is based on marketing theory. Zeithaml and Zeithaml's [24] marketing theory also applies here because it provides continuous guidance for game development organizations on how they should react to opportunities and how, by taking appropriate market actions, the organization can create opportunities by changing the environment. Hunt [25] describes marketing as a management responsibility that helps in sensing the market and articulating new and valuable propositions. Berry [26] also highlighted the use of customer relationship management (CRM) to develop an appropriate marketing strategy to retain, attract, and enhance customer relationships. Gronroos [27] defined marketing in the context of CRM, and Fornell and Wernerfelt [28] described a marketing strategy aimed at attracting new and retaining existing customers and resulting in increased revenue and profitability. Owomoyela *et al*. [29] described how organizations can develop their marketing strategies in a way that enables them to build, maintain, and defend their competitive advantage. Managerial judgment will be helpful in identifying strategic marketing uncertainties and environmental ambiguities.

In the literature, very few studies have described market orientation for the DGI. Lee *et al*. [30] suggested that game developers must develop market reports in the requirements engineering phase and during game distribution. The marketing team plays an important role in this. The main activities performed by marketing teams along with the CRM team are packaging, advertising, management of marketing agents, and production of a complete marketing plan. Kasaliaki and Mustafee [31] explored sustainable development strategies for a serious game audience. Analyzing the characteristics and requirements of the target audience helps developers generate a sustainable game development process. Xin [32] highlighted the barriers in serious mobile game markets and the current market segmentation for serious games. Before developing serious games, developers must analyze the market segment and their own competitive advantage. This study highlights the issue of market analysis before starting a game project to determine what types of games are in demand.

## 2.3 Innovation

Especially in the DGI, innovation has a special place as a key driver of economic growth and competitiveness. Innovation has many forms and has become known as a critical dimension of achieving better economic performance, especially in knowledge-driven economies. Innovation can be defined as the successful exploitation of new social or commercial ideas and the ability, once new ideas have been brought to market, to reduce cost, improve services, and improve existing arrangements by offering new and effective alternatives. Afuah [33] defined strategic innovation for organizations as follows: "a strategic innovation is a game-changing innovation in products/services, business models, business processes, and/or positioning of competitors to improve performance". Johannessen [34] described a systematic perspective on innovation theory. He considered 14 propositions from the literature and investigated the connection between economic crises and innovation. He categorized innovation into two major categories: institutional and economic innovation. Furthermore, institutional innovations were categorized into political, cultural, and social innovations. The economic innovations category consisted of organizational, material, service and market innovations. Basically, innovations in organizations are associated with managing an organization in new ways as well as with new business models. A business model innovation framework has been proposed by Comviva Technologies [35] that contains an industry model (adoption of new industries by redefining existing ones), a revenue model (reconfiguration of offerings and a pricing model), and an expertise model (value-chain role playing). Lindgardth *et al.* [36] also proposed an innovative business model including two elements: a value proposition and an operating model. The value proposition is about who the target audience is, what kind of product/service the organization will offer, and what the organizational revenue model will be. The operating model addresses the issue of service/product delivery that generates profitability and includes three critical areas: the value chain, a cost model to generate revenue, and an organization that develops and deploys assets to enhance and sustain competitive advantage.

## 2.4 Relationship Management

Effective CRM is a highly critical element in the success of any business. Wilson [37] observed that relationship management basically involves developing and maintaining long-term, close, satisfying and mutually beneficial relationships between customers and organizations based on collaboration and trust. In relationship management, customer profiling, promotional strategies, customer service and support, customer information, organizational behavior, and channel management are all contributing factors. Recently, organizations have been integrating their customers into the design, production, or delivery of goods and services. These organizations are mainly targeting revenue increases or cost reductions by relying on their customers as co-producers of goods or services that they offer to the market. This trend towards integrating users or customers shows that new organizational choices are being made by companies to generate high margins. This is a fundamental change in business strategy that pushes organizations to think about new ways to mobilize their users to increase revenue. Plé *et al.* [38] explained the role of customers in this business model. They proposed a theoretical framework called the Customer-Integrated Business Model (CIBM) by combining customer participation with the business model literature. The framework based on ROCA (Resource Oriented Client Architecture) and proposed by Lecocq *et al.* [39] considered the customer as a resource; the model was illustrated by two case studies. They concluded that more field research is required to explore the relationship between the customer-as-a-resource approach and business profits. Most studies in the literature consider customer participation in service marketing and management, whereas only a few consider customer integration as a resource. The digital game literature also lacks the dimension of customer integration for business performance. Stanely *et al.* [40] looked at user integration from a different perspective. They described the cumulative context of a digital game and accumulated all contextual information on a player's activity using mobile sensors to change the game state. Experimental results indicated that the player found the game engaging and fun. Ermi and Mayra [41] pointed out that user involvement is a multi-dimensional and complex phenomenon which is not totally dependent on the nature of the specific genre or game, but also upon each player's choices or preferences.

## 2.5 Time to Market

The time-to-market phenomenon has long been recognized as a crucial enabler for business success. From this perspective, organizations can be categorized into pioneers, early followers, and late movers [42, 43]. The pioneers emerge as solution providers in the market and gain a sustainable competitive advantage over followers. This enables them to amass a major part of the market, making it more difficult for successors to gain market share. Hence, the timing of entry into the market becomes more crucial for organizations to gain profit and competitive advantage. Products that enter the market at the right time or have short time to market have potential of success. A digital game organization's ability to reach the market before their competitors and gain adoption is an important factor in the long-term success of games. The time-to-market process in the DGI can be defined as integration of new technology into digital game production. Today, digital game organizations can gain competitive advantage by introducing the next generation of technologies into the game market through new game development strategies that enable them to be first in the game industry market. Very few studies in the literature have highlighted the importance of the time-to market factor specifically in the digital game industry [44,30], and none of them has discussed it from a business performance perspective.

## 2.6 Monetization Strategy

The DGI sector is learning the game of monetization. Around the world, millions of consumers play games on either online media portals or social networking sites every month. Monetization strategy is very important because it is a risky business. It provides an insight into the organization of a business that either is worthwhile or is not. Monetization strategy in games is similar to the setting of financial objectives for any organization. Financial objectives are defined as organizations set their financial targets over a certain period of time. Financial objectives are different from other types of organizational objectives such as business or customer retention objectives because they cannot be easily measured monetarily if achieved. Game development involves high costs, and only the top 5% of games in the market are profitable. A game that fails in the market can lead to severe losses or even bankruptcy in the case of small developers. The organization needs therefore to have proper financial management and appropriate financial planning to ensure that enough funding is available when needed. Second, financial controls determine whether the organization is meeting its financial objectives. Finally, financial decision-making is itself very important [45].

In social games, players are able to create their own virtual characters and communities and interact with their friends. Companies involved in the game business have developed business models for paid content such as subscription, advertising, and micro-transactions for virtual goods. In general, users are not interested in paying for virtual goods, but the few who pay for them make this business model work. Eventually, micro-transactions, especially in the social game lifecycle, have become a driver for incremental revenue. In the massive multiplayer online game sector (MMO), the bulk of game revenue is still generated by subscriptions, but use of micro-transactions is growing for virtual goods. The importance of a monetization strategy for the DGI has been explored by only two studies, but not in detail, and neither of them discussed its impact on business performance [46].

## 2.7 Brand Name Strategy

A brand name is regarded as a crucial enabler for business success in any organization. The brand is considered as both a point of comparison with other products and a promise of quality to the customer. Bennett [47] described a brand as a term, name, symbol, sign, design, or combination of any of these concepts that helps to identify the products or services of a particular seller. Generally, the brand name has high impact on the organization's business. Between the organization and its customers, branded products serve as an interface, and brand loyalty enables marketing by word of mouth. The organization's brand name strategy has a strong impact on the customer decision-making process. Bergstrom [48] perceived that in the case of products and competitors that are easily replicable or duplicated, brands help customers in the decision process of buying a particular product.

Hence, the DGI has successfully adopted a brand name strategy in the game development process. In games, there are many successful platform brands, including Nintendo, Sony, and Microsoft for consoles, Apple (IOS), Samsung (Android), and others for mobile platforms, and Windows, Apple, and others for PCs. However, no study has described the brand name strategy in the game development process and its impact on business performance.

## 3. Research Model and Hypotheses

The main objective of the proposed research model is to analyze the interrelationship between key factors and game business performance and also to understand the influence of these factors on a game development organization's business performance in the DGI market. Davenport [49] and Aguilar-Sav'en [50] described the combination of structured business process activities in an organization to achieve specific goals. The model's theoretical foundation is based on a combination of existing concepts found in the game development literature and business models for the game industry. It is worth noting that most studies in the literature discuss one or two of the factors mentioned above in the context of game development organizations and their impact on game performance. To the best of the authors' knowledge, this is the first study in the game development literature that highlights key factors for game performance in game development organizations. This study proposes to investigate empirically the influence and association of key factors in game development organizations and game business performance. The Fig. 1 presented the theoretical research model of this study to be empirically investigated. The theoretical model evaluates the relationships of different independent variables emerging from organizational concepts such as organizational management, theory, and behavior in the context of game development organizations on the dependent variable of game business performance within the organization. This study mainly investigates and addresses the following research question:

**Research Question:** What is the impact of key business factors on overall game business performance in the DGI?

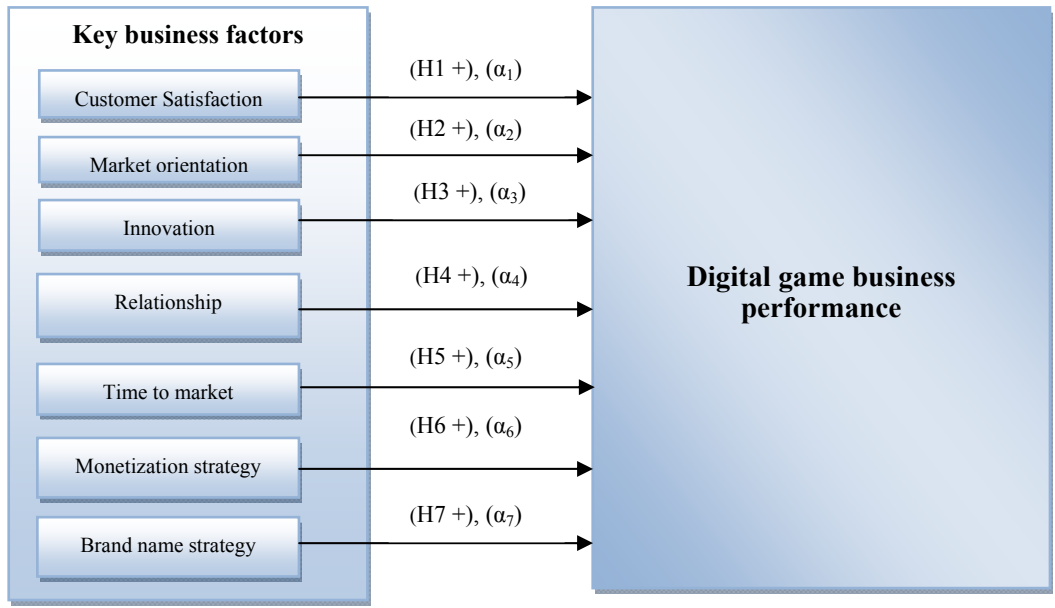

Fig 1. Research model

The research model includes seven independent variables: customer satisfaction, market orientation, innovation, relationship management, time to market, monetization strategy, and brand name strategy, and one dependent variable: the business performance of the digital game. The multiple linear regression equation of the model is given as Equation 1:

Business performance of game = $\beta_0 + \beta_1 f_1 + \beta_2 f_2 + \beta_3 f_3 + \beta_4 f_4 + \beta_5 f_5 + \beta_6 f_6 + \beta_7 f_7$,        (1) where $\beta_0$, $\beta_1$, $\beta_2$, $\beta_3$, $\beta_4$, $\beta_5$, $\beta_6$, $\beta_7$ are coefficients and $f_1$-$f_7$ are the seven independent variables. For purposes of empirical investigation, the following hypotheses are stated:

**H1:** Customer satisfaction has a positive impact on the performance of a digital game.
**H2:** Market orientation has a positive impact on the performance of a digital game.
**H3:** Innovation has a positive impact on the performance of a digital game.
**H4:** Relationship management has a positive impact on the performance of a digital game.
**H5:** Time to market has a positive impact on the performance of a digital game.
**H6:** Monetization strategy has a positive impact on the performance of a digital game.
**H7:** Brand name strategy has a positive impact on the performance of a digital game

## 4. Research Methodology

Digital game development organizations are involved in various business activities such as game development, publishing, distribution, and finally customer engagement. The targeted respondents of this study were employees of game development organizations or independent studios. Some organizations handled all these activities by themselves, whereas some of them outsourced publishing or distribution activities. Initially, the authors joined various game development community forums and started blogs about a data collection request for an empirical study. A survey questionnaire was also created using the Survey Monkey Web site, and personalized emails were sent to various organizations. The respondents were from multinational companies in Asia, Europe, and North America, and statistics about them are illustrated in Fig. 2. Participant organizations agreed to take part in the study based on mutual agreement that their identities would be kept confidential. The size of the participating organizations varied from micro to large scale. Micro size organizations consisted of 3–5, small ones of 5–99, medium ones of 100–499, and large ones of 500+ team members belonging to various departments within the organization. Figure 3 show the number of respondents by organization size.

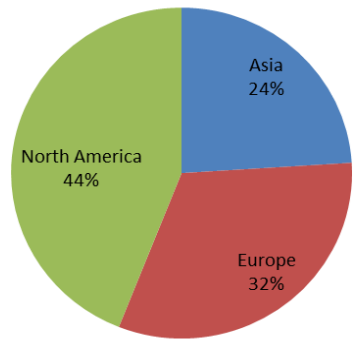

Fig. 2 Number of respondents by continent

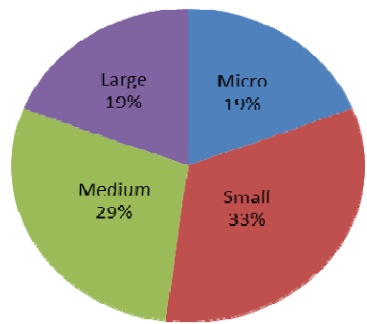

Fig. 3 Number of respondents by organization size

The participating organizations mainly developed games for different platforms such as kiosks and standalone devices, the Web, social networks, consoles, PC/Macs, and mobile phones. The game genres implemented in most of their projects included action or adventure, racing, puzzles, strategy/role playing, sports, music-based, and other categories. The participant organizations distributed the surveys within various departments, and the survey respondents had been employed in that particular organization for at least three years. The survey respondents worked in various capacities from game development to middle and senior management and played a role in either policy-making or implementation of organizational strategies. The total number of survey respondents was 61, including a minimum of two and a maximum of five responses from each organization.

*4.1 Measuring instrument*

This study gathered data on the key business factors and the perceived level of game performance identified in the research model depicted in Fig. 1. To learn about these two topics, the questionnaire presented in Appendix A was used as a data collection instrument. First, organizations involved in the game development business were asked to what extent they practiced the identified key business factors within their organization. Second, they were asked about the business performance of their games in the digital game industry. The five-point Likert scale was used in the questionnaire, and with each statement, the respondents were required to specify their level of agreement or disagreement. Thirty-three items were used to measure the independent variables (the key factors), and for the dependent variable (game performance), eight items were used. The literature related to key business factors was reviewed in detail to ensure a comprehensive list of measuring items for each factor from the literature. A multi-item, five-point Likert scale was used to measure the extent to which each key business factor was practiced within the organization. The Likert scale ranged from (1) meaning "strongly disagree" to (5) meaning "strongly agree" and was associated with each item. The items for each identified factor were numbered 1 to 33 in Appendix A and also labelled sequentially. Game business performance was the dependent variable and was measured for at least the past three years in the context of organizational financial strength, market growth, cost savings, and reduced development time based on a multi-item five-point Likert scale. The designated items for the dependent variable were numbered separately from one to eight and labelled sequentially. All the items specifically written for this study are presented in Appendix A. To the best of the authors' knowledge, this is the first empirical study of key digital game business factors in the DGI.

*4.2 Reliability and validity analysis*

The empirical studies included two integral measures of precision: reliability and validity. Reliability refers to a measurement's reproducibility or consistency, whereas validity refers to the inference or agreement between the true value and the measured value. A reliability and validity analysis was performed for the measuring instrument that was specifically designed for this empirical study. This analysis was based on the most common approaches used in empirical studies. Based on internal consistency analysis, reliability of the multi-scale measurement items for the seven identified factors was evaluated. Cronbach's alpha coefficient [51] was used to evaluate internal consistency. First, Cronbach's alpha coefficient was calculated on a sample dataset which excluded assessment items from each category if they affected the desired value of Cronbach's alpha coefficient. The responses to question 6 on market orientation, question 12 on innovation, question 22 on time to market, question 24 on monetization strategy, and question 33 from brand name strategy were excluded from the investigation based on their effect on Cronbach's alpha coefficient. The reliability analysis for the seven factors is reported in Table 1, with Cronbach's alpha coefficient ranging from 0.61 to 0.76. Satisfactory value criteria for Cronbach's alpha have been reported by a number of researchers based on their findings. Nunnally and Brenste [52] stated that a value of the reliability coefficient of 0.70 or higher for a measuring instrument was satisfactory. Van de Ven and Ferry [53] suggested that a value of 0.55 or higher of the reliability coefficient could be considered satisfactory. Osterhof [54] concluded based on his findings that a value of 0.60 or higher was satisfactory. Hence, all developed variable items for this study could be considered reliable.

Campbell and Fiske [55] concluded that convergent validity has occurred if scale items are highly correlated and in a given assembly, they move in the same direction. For the analysis of convergent validity, principal component analysis (PCA) [56] with seven factors was used, with the results shown in Table 1. The eigenvalues [57] were used as a reference point to determine the construct validity of the PCA-based measuring instrument. For this empirical investigation, the Kaiser

Criterion [58] was used, which states that any value greater than one for any component is to be retained. The eigenvalue analysis showed that out of the seven variables, six together formed a single factor, whereas brand name strategy loaded on two factors, and both eigenvalues were greater than one. The reported convergent validity of this study has been considered as adequate.

Table 1
Cronbach's alpha coefficient and principal component analysis of seven variables

| Business factor | Item no. | Coefficient α | PC eigenvalue |
|---|---|---|---|
| Customer satisfaction | 1–4 | 0.71 | 1.49 |
| Market orientation | 5–8 | 0.67 (Q6) | 1.57 |
| Innovation | 9–13 | 0.74 (Q12) | 1.01 |
| Relationship management | 14–19 | 0.60 | 1.16 |
| Time to market | 20–23 | 0.64 (Q22) | 1.61 |
| Monetization strategy | 24–28 | 0.61 (Q24) | 1.25 |
| Brand name strategy | 29–33 | 0.76 (Q33) | 1.79 |

### 4.3 Inter-rater agreement analysis

Mostly, there were two or one respondents from one organization. We have performed inter-rater agreement analysis [59] to address the issue of conflicting opinion from same organization. The inter-rater agreement is about the level of agreement in the ratings provided by different respondents for the same process or software engineering practice [60]. Thus, we performed inter-rate agreement analysis to identify the level of agreement among different respondents from same organization. To evaluate inter-rater agreement, Kendall co-efficient of concordance (W) [61] is usually preferred for ordinal data as compared others methods like Cohen's Kappa [62]. "W" represents the difference between the actual agreement drawn from data and perfect agreement. Values of Fleiss Kappa and the Kendall's $W$ coefficient can range from 0 (represents complete disagreement) to 1 (represents perfect agreement) [63]. Therefore, Kappa [60] standard includes four levels ranging from: $< 0.44$ means poor agreement, 0.44 to 0.62 entails moderate agreement, 0.62 to 0.78 indicates substantial agreement, and $> 0.78$ represents excellent agreement. In this study, the Kappa coefficient observed come under substantial category and ranges from 0.63 to 0.68. Table 2 reports the Kappa and Kendall statistics for five organizations.

Table 2
Inter-rater agreement analysis

| Organizations | Kendall's statistics | | Kappa Statistics | |
|---|---|---|---|---|
| | Kendall's Coefficient of Concordance (W) | χ2 | Fleiss Kappa Coefficient | Z |
| A | 0.72 | 58.20* | 0.68 | 8.20* |
| B | 0.65 | 52.90* | 0.63 | 7.98** |
| C | 0.71 | 57.42* | 0.67 | 8.04* |
| D | 0.63 | 51.32* | 0.62 | 7.54* |
| E | 0.74 | 60.14** | 0.69 | 9.01** |

*Significant at $p<0.05$                    **Insignificant at $p>0.05$

### 4.4 Data analysis techniques

Various statistical approaches were used in this research for data analysis. Initially, this activity was split into three phases to estimate the significance of hypotheses H1–H7. Phase I involved parametric statistics and normal distribution tests. In Phase II, partial least squares (PLS) was used as a nonparametric statistical approach. Due to the small sample size, both parametric and nonparametric approaches were used to address the threat to external validity. Multiple items were used in the measuring instrument for each independent variable and the dependent variable, with respondents' ratings for each variable aggregated to obtain a composite value. In phase I, tests were conducted for each hypothesis H1–H7 using parametric statistics such as the Pearson correlation coefficient and the one-tailed $t$-test. In phase II, nonparametric statistics such as the Spearman correlation coefficient were used to test hypotheses H1–H7. In phase III, tests were carried out for research model hypotheses H1–H7 based on the PLS technique. Fornell and Bookstein [64] and Joreskog and Wold [65] reported that if non-normal distribution, complexity, small sample size, and low theoretical information are issues, then PLS will be helpful. The PLS technique was used in Phase III to increase the reliability of the results and deal with the limitation of small sample size. The main reason for the small sample size was first, that most games on the market are developed by one or three developers, but this study targeted game development companies with more than three employees, and second, some companies declined to respond to the survey due to their busy schedule. For statistical calculations, the Minitab 17 software was used.

## 5. Data Analysis and Results

### 5.1 Phase I of hypothesis testing

Parametric statistics were used in this phase to test hypotheses H1–H7. The Pearson correlation coefficient was examined between the independent variables (key business factors) and the dependent variable (game performance) of the research model, as illustrated in Figure 1. To accept a hypothesis, the level of significance was selected so that if the $p$-value was less than 0.05, the hypothesis would be accepted, and if the $p$-value was greater than 0.05, the hypothesis would be rejected [66]. The calculated results for the Pearson correlation coefficient are listed in Table 3. Hypothesis H1 was accepted because the Pearson correlation coefficient for customer satisfaction and game performance was positive (0.50) at $p<0.05$. For hypothesis H2 concerning market orientation and game performance, the Pearson correlation coefficient was also positive (0.57) at $p<0.5$, and therefore hypothesis H2 was also accepted. Hypothesis H3 concerning innovation and game performance was rejected due to its higher $p$-value (0.93). Hypothesis H4 concerning relationship management and game performance was also rejected based on its negative Pearson correlation coefficient (-0.361) at $p<0.05$. Hypothesis H5 concerning time to market and game performance was accepted based on its positive correlation coefficient (0.61) at $p<0.05$. Hypothesis H6 regarding monetization strategy and game performance was also accepted due to its positive Pearson correlation coefficient (0.25) at $p<0.05$. The last hypothesis (H7) between brand name strategy and game performance was also found to be significant (0.79) at $p<0.05$ and was therefore accepted. Hence, in summary, hypotheses H1, H2, H5, H6, and H7 were accepted and found to be statistically significant. Hypotheses H3 and H4 were not supported statistically and were therefore rejected.

Table 3
Hypothesis testing using parametric and nonparametric correlation coefficients.

| Hypothesis | Key factor | Pearson correlation coefficient | Spearman correlation coefficient |
|---|---|---|---|
| **H1** | Customer satisfaction | 0.50* | 0.55* |
| **H2** | Market orientation | 0.57* | 0.57* |
| **H3** | Innovation | 0.01** | 0.13** |
| **H4** | Relationship management | -0.16** | -0.16** |
| **H5** | Time to market | 0.61* | 0.55* |
| **H6** | Monetization strategy | 0.25* | 0.27* |
| **H7** | Brand name strategy | 0.79* | 0.78* |

*Significant at $p<0.05$                    **Insignificant at $p>0.05$

### 5.2 Phase II of hypothesis testing

Phase II involved testing hypotheses H1–H7 based on the nonparametric Spearman correlation coefficient. The observations made in this phase for the Spearman correlation coefficient are also reported in Table 2. Hypotheses H1 was accepted because of its positive Spearman correlation coefficient (0.55) at $p<0.05$. The Spearman correlation coefficient for market orientation and game performance (hypothesis H2) was also positive (0.57) at $p<0.05$ and was also found to be significant. The relationship between innovation and game performance (hypothesis H3) was not found to be statistically significant due to its Spearman correlation coefficient (0.13) at $p>0.05$ and was rejected. For hypothesis H4, the Spearman correlation coefficient was negative at $p<0.05$, and therefore H4 was rejected. Hypothesis H5 concerning time to market and game performance was accepted due to its positive coefficient (0.55) at $p<0.05$. Hypothesis H6 concerning monetization strategy and game performance was also accepted due to its positive Spearman correlation coefficient (0.27) at $p<0.05$. The last hypothesis (H7) between brand name strategy and game performance was also found to be significant (0.78) at $p<0.0$. Hence, in summary, hypotheses H1, H2, H5, H6, and H7 were accepted and found to be statistically significant. Hypotheses H3 and H4 were not supported statistically and were therefore rejected.

### 5.3 Phase III of hypothesis testing

Phase III included hypothesis testing based on the partial least squares (PLS) technique. PLS was used for cross-validation and to overcome some limitations associated with the results obtained from the parametric and nonparametric statistical approaches used in Phases I and II. Hypotheses H1–H7 were tested for direction and significance. To examine PLS for each hypothesis, the dependent variable (game performance) was designated as the response variable and the individual business factors as the predicate variable. The observed structural test results for the hypotheses are reported in Table 4 and include the observed values of $R^2$, the path coefficient, and the $F$-ratio. The path coefficient for customer satisfaction (H1) was observed to be 0.78, $R^2$ was 0.24, and the $F$-ratio was 19.10, and H1 was found to be significant at $p<0.05$. Market orientation (H2) had a positive path coefficient of 1.04, $R^2 = 0.32$, and $F$-ratio $= 28.51$ and was also found to be statistically significant at $p<0.05$. Innovation (H3) had a path coefficient of 0.02, a very low $R^2$ of 0.0001, and an $F$-ratio of 0.0001 and was found to be insignificant at $p<0.05$. Relationship management (H4) had a negative path coefficient of -0.27, a low $R^2$ of 0.01, and an $F$-ratio of 1.69 and was judged to be insignificant because the $p$-value was greater than 0.05. Time to market (H5) (path coefficient: 1.16, $R^2$: 0.37, and $F$-ratio: 35.52) had the same direction as proposed. Monetization strategy (H6) (path coefficient: 0.51, $R^2$: 0.64, and $F$-ratio: 4.04) and brand name strategy (path coefficient: 0.94, $R^2$: 0.62, and $F$-ratio: 100.38) was found significant at $p<0.05$.

Table 4
PLS regression results for hypothesis testing

| Hypothesis | Factors | Path coefficient | $R^2$ | $F$-Ratio |
|---|---|---|---|---|
| H1 | Customer satisfaction | 0.78 | 0.24 | 19.10* |
| H2 | Market orientation | 1.04 | 0.32 | 28.51* |
| H3 | Innovation | 0.02 | 0.01 | 0.01** |
| H4 | Relationship management | -0.27 | 0.01 | 1.69** |
| H5 | Time to market | 1.16 | 0.37 | 35.52* |
| H6 | Monetization strategy | 0.51 | 0.64 | 4.04* |
| H7 | Brand name strategy | 0.94 | 0.62 | 100.38* |

*Significant at $p<0.05$                    **Insignificant at $p>0.05$

### 5.4 Research model testing

The linear regression equation for the research model is given by Equation 1. The research model was tested to provide empirical evidence that business factors play a considerable role in digital game performance in the market. The test procedure examined the regression analysis, the model coefficient values, and the direction of the associations. The dependent variable (game performance) was designated as the response variable and the other independent variables (all the key business factors) as predicate variables. The regression analysis model results are reported in Table 5. The path coefficients of five of the seven variables (customer satisfaction, market orientation, time to market, monetization strategy, and brand name strategy) were positive and found to be statistically significant at $p<0.05$. The path coefficient of innovation was positive, but was found not to be statistically significant at $p<0.05$. The path coefficient of relationship management was negative and made this factor insignificant in the research model. The overall $R^2$ value of the research model was 0.74, and the adjusted $R^2$ value was 0.71 with an $F$-ratio of 21.16, which was significant at $p<0.05$.

Table 5
Linear regression analysis of the research model

| Model coefficient name | Model coefficient | Coefficient value | t-value |
|---|---|---|---|
| Customer satisfaction | $\beta_1$ | 0.23 | 1.67* |
| Market orientation | $\beta_2$ | 0.66 | 3.35* |
| Innovation | $\beta_3$ | 0.18 | 1.20** |
| Relationship management | $\beta_4$ | -0.14 | -1.05** |
| Time to market | $\beta_5$ | 0.02 | 1.10* |
| Monetization strategy | $\beta_6$ | 0.13 | 1.68* |
| Brand name strategy | $\beta_7$ | 0.69 | 4.13* |
| Constant | $\beta_0$ | 7.35 | 2.01* |
| $R^2$ | 0.74 | Adjusted $R^2$ | 0.71 |
| $F$-ratio | 21.16* | | |

*Significant at $p<0.05$          **Insignificant at $p>0.05$

## 6. Discussion

Today's digital era has attracted many people to play games and to develop their own games for profit. Developing a digital game involves activities from different disciplines and has its roots in business management and software engineering. This research aims to help game development organizations understand the interdependencies and relationships between key business factors and game performance in the market. This research offers an opportunity to explore empirically the association between key business factors and digital game performance. This is the first empirical investigation of business factors in relation to game performance, and the results support the theoretical foundations and provide first evidence that key business factors play an important role in digital game performance. This could well result in institutionalizing the digital game production approach in the game development organization, which in turn has a high potential to maximize profits.

Customer satisfaction in the DGI refers to meeting the customer's expectations by providing a functional game, addressing the availability issue for online games, and offering good customer service and expert advice on games. The customer satisfaction variable for business performance measurement in the DGI has not yet been explored in the literature. Basically, game development organizations must value their customers or players by meeting their expectations. This study has found a positive association between customer satisfaction and digital game performance. Organizations can use appropriate measures to track their customers' purchasing behavior and focus more on providing customer service. To implement better customer service, organizations need to understand their game players, implement player-specific platform services, and take feedback strongly into consideration. Most literature reviews have focused on the relationship between business performance and customer satisfaction in different industries. To be successful in the competitive DGI market, game development organizations must take all these strategies into account to explore their relationship with their customers. By adopting best practices, organizations will be able to understand their customers or players, and instead of aiming for one-hit wonders, attracting new customers and retaining existing ones will become the main indicators of customer satisfaction. Important factors affecting customer retention include their initial play experience, the level of game addictiveness, the fit between organizational targets and the market, and finally, the ability of the organization to correct all issue that harm retention. Customer satisfaction data in an organization are also helpful for continuous improvement, which affects the organization's business performance on a long-term basis.

Market orientation was also found to have a positive impact on digital game performance. In the DGI, market orientation is a vast and complex topic. Game development organizations need to focus mainly on two artifacts while developing their games. First, the marketing strategy artifact is a kind of guideline that describes your targets, and second, the marketing plan artifact is a detailed description of your targets and how you will execute them. The organization must develop the marketing strategy at the beginning of the game development process because most of the decisions about game development such as monetization, game design, languages, and demographic locations of game availability will impact the marketing strategy. For market-driven games, one important decision about marketing is whether the organization will publish the game by itself or transfer it to a publisher. In each case, the marketing plan execution will be different. A publisher will take into account the target audience, locations, and platforms, and the marketing plan will be executed by the publisher. However, if the organization publishes the game on their own, its must also consider its target audience, the game business model, geography, budget, platforms, and marketing channels. The impact of market orientation on business performance was explored by Adewale *et al.* [67], who reported that market orientation is a significant joint predicator of business performance in terms of return on investment, market share, and profitability. Business performance as a market orientation variable can be measured in terms of monetization, packaging, and promotion strategies as well as calculations of individual customer revenue and profitability.

The DGI appears to consider innovation as a basic source of competiveness. Most organizations see innovation in games as bringing new things to the market and being different from competitors. Innovation in game development can involve application of new ideas at the game level, storyboard production, use of new technology, or the creative artistry of the game, with the aim of attracting more gamers and thus creating value in terms of business performance. Not one single study has addressed the issue of innovation in the DGI. Innovation in the game industry can also refer to an innovative business model of the game development process that addresses all innovation categories, as described by Johannessen [34] and Lawson and Samson [68]. On contrary, the findings of this empirical investigation do not support a statistically positive relation between innovation and digital game performance. The direction of association was positive, but the required statistical level of confidence was not supported.

It has also been assumed here that the user integration approach in the DGI enables organizations to use their users as a resource. It is important to consider users as a resource because especially in the computer game industry, users are the revenue producer, and the business totally depends on their positive playing experience. More user involvement enables the organization to retain its users/customers. The question now arises of how game users who are also players can become involved in parts of the game development process. One way of user integration is through virtual community membership. Nohria and Ghoshal [69] argue that "the real leverage lies in creating a shared context and common purpose and in enhancing the communication densities within and across the organization's internal and external boundaries". This argument also supports the concepts of customer socialization and community participation in the game development process. However, this user integration approach is cost-effective for any organization. In such communities, customers can participate based on their broad communities of interest. They can be a part of game development by sharing their playing experiences, being involved in idea generation, becoming co-creators or testers of games, or in other ways. Use of online communities in the development process constitutes an important source of innovation and also enables organizations to implement constructive relationships with their users.

In successful game development, relationship management plays a significant role. Integrating players into the development process and maintaining excellent working relationships with them helps developers to improve the performance and functionalities of their games. However, the assumption that relationship management also helps the organization to understand its customers' needs better and remain up-to-date about market trends was not found to be significant in this study. Empirical investigation found a negative association and also insufficient statistical support for a significant confidence level. Hence, the study was not able to find any impact or association between relationship management and digital game performance.

Because the DGI is flourishing, competition is very tough between digital game organizations. The organization which achieves competitive advantage using time-to-market processes will have a positive impact on business performance. This hypothesis was also supported by empirical investigation. Hence, game launch timing is important to capture major market share. The time-to-market approach in a game development organization develops a publishing schedule for the game and provides essential guidelines for development schedules to the developers. The game launch schedule is a crucial business decision that has profound and long-lasting impact on the business performance of an organization in retaining and capturing the market.

In the DGI, fulfillment of financial objectives or monetization strategy depends on economically optimizing the pricing scheme for customers, the cost structure, and the target customer segment. In this empirical investigation, a positive association was found between monetization strategy and digital game performance. The pricing scheme can be a one-time payment, pay per session, pay per play, or subscription-based or bundled pricing. The cost structure is based on the complete picture of the entire budget for game development, including marketing and distribution costs. The overall cost of each phase until delivery to the user directly impacts the overall profitability of the organization. However, it is difficult to measure the cost impact of each phase on overall business performance [46]. In this situation, the impact of monetization can be measured by using the overall profitability of the organization as a measure of business performance. As for the target customer segment, it is important to understand the needs of target customer groups to ensure that games are properly priced, marketed, and packaged to achieve business success.

Recently, in the DGI, use of successful game development brands that are useful to particular market segments has helped organizations connect with their target audience. This empirical investigation found a positive association and impact of brand name strategy on overall game performance. In particular, brand name strategy has become marketing strategy in branded games. Although branded game development costs the organization more, it pays off after publication by attracting large numbers of new and repeat users. An effective brand name strategy helps in defining game development and execution, ensuring that the game gets appropriate promotion in the marketplace, and positioning the game for its target audience.

Overall, the findings of the study are important for the development of good quality digital game. Rapid and continual changes in technology and intense competition not only affect the business, but also have a great impact on development activities. To deal with this strong competition and high pressure, game development organizations must continuously assess their activities and adopt a proper evaluation methodology. Use of a proper assessment methodology will help the organization identify its strengths and weaknesses and provide guidance for improvement. However, the fragmented nature of the game development process requires a comprehensive evaluation strategy which has not yet been entirely explored. The findings of the study will help the game development organizations to look for contributing key success factors from business perspective. This study is a part of a larger project aiming to propose digital game maturity assessment model. Business perspective is one of the identified dimensions out of developer, consumer and process itself. The findings of this study also provide the justification to include these factors in the process of assessment methodology.

### 6.1 Limitations and Threats to External Validity

Metrics, surveys, case studies, and experiments are some examples of empirical techniques used for software engineering processes and product investigations. However, certain limitations are associated with these empirical investigations and with this study as well.

The first limitation of this study was the selection and choice of independent variables. Seven independent variables were included to investigate their association with and impact on digital game performance. However, other key factors may exist that have a positive impact on digital game performance, but this study was limited to the seven variables because of their presence in the literature. In addition, other key factors may exist, such as environmentally based, regionally based, or political factors, which have a positive impact on digital game performance, but are not considered in this study. Furthermore, this study has focused only on business factors in digital game performance.

The second notable limitation of the study is the small sample size. Most game industry developers who follow either agile practices or poor development practices were unable to respond to the questionnaire and did not respond. The vast majority of game developers work in one- to three-person teams and did not have the required level of experience (three years) and were therefore excluded from this empirical investigation. Most respondents refused to answer the questionnaire because they were too busy in the game development process or launching their games in the market. Some game development organizations are also hesitant to disclose their business performance. Therefore, data collection from the game industry was limited, resulting in small sample size. The number of respondents from one organization was beyond the authors' control because the organization's upper management was responsible for distributing the survey within a company. The main effect of small sample size is on its statistical power, Type II error, significance and on distribution [70]. Therefore, the important thing is while making conclusion avoid strong statements. As, the small samples size studies results can be difficult to replicate or generalize [71] but they do provide some interplay between variables. The well designed small studies are seems ok to conduct as they provide quick results but they need to be interpreted carefully [72]. The low sample size constraint of this study makes the results difficult to generalize. However, the results of this study are useful in providing some basic foundation to design larger confirmatory study, which is the future objective of this work.

Biased decision-making was the third limitation of this study. Although multiple responses were collected from each company to address the bias issue, but it remained a core issue. Respondents were asked to consult available documentation within a company to fill out the survey. Accepted psychometric principles were used to design the assessment items, for conflicting opinions from same organization inter-rater agreement analysis was performed but the measuring instrument was still based on individual subjective assessment.

In spite of its specific and general limitations, this study has contributed to the field of digital games and has helped game development organizations to understand the business dimension of digital games.

## 7. Conclusions

Game development is an interdisciplinary concept that embraces software engineering, business, management, and artistic disciplines. This research has facilitated a better understanding of the business dimension of digital games. The main objective of this research was to investigate empirically the effect of business factors on the performance of digital games in the market and to try to find answers to the research questions posed in this study. Empirical investigation results demonstrated that business factors play an important role in digital game performance. The results of the study strongly indicate that customer satisfaction, time to market, monetization strategy, market orientation, and brand name strategy are positively associated with the performance of a digital game organization. The empirical investigation found no strong association or impact between relationship management or innovation and digital game performance.

This study is the first of its kind in the field of digital games. It will help and enable organizations to achieve a better understanding of the effectiveness of business factors and their role in terms of game performance in the market. Game development organizations need to consider these various business factors over and above their current efforts to improve the performance of their developed games in the market.

Currently, the authors are working on developing a digital game maturity model for game development process assessment. This study has provided the empirical evidence and justification to include business factors in evaluating the business dimension of game development process maturity.

## Appendix A: Key business factors measuring instrument

### *SECTION ONE*
### *1.1 Participant details*

| Full Name (Optional) | | Job Title/Position | |
|---|---|---|---|
| Experience (in years) | | | |
| Address | | | |
| Telephone no. (optional) | | | |
| Email | | | |

### *1.2 Demographics*

| Country in which the company is located? |
|---|
| *Please Specify:* |

| What is the scope of your company? |
|---|
| National ☐    Multinational ☐    Don't Know ☐ |
| *Please Specify:* |

| Approximately how many people are employed by your company? ( Please tick the appropriate box) |
|---|

| Less than 20 ☐ | 20-70 ☐ | More than 100 ☐ | Not sure ☐ |
|---|---|---|---|

*Please Specify:*

| What type of game genre is developed by your company and what is the target platform for developed games? |
|---|
| *Please Specify:* |

| Who are the target audience? |
|---|
| *Please Specify:* |

## *SECTION TWO*

### 2.1   Evaluation of business performance success factors identified through literature review

The questionnaire objective is to find out which factors have a positive impact on business performance. Please select the correct scale based on your best knowledge.

| Business performance key factors for game development companies | | | | | | | |
|---|---|---|---|---|---|---|---|
| Likert scale (1 = strongly disagree; 2= disagree; 3 = neutral; 4= agree; 5 strongly agree) | | 1 | 2 | 3 | 4 | 5 | N/A |
| **Customer or player satisfaction** | | | | | | | |
| 1 | The organization is using a game rating scheme to respond to player requests. | | | | | | |
| 2 | The organization has a good customer service department. | | | | | | |
| 3 | The organization provides expert advice on games. | | | | | | |
| 4 | The organization provides feedback or response to its customers. | | | | | | |
| **Market orientation** | | | | | | | |
| 5 | The organization has adequate skills and resource to perform detailed market studies to determine what types of games are in demand and who will be the target audience. | | | | | | |
| 6 | The organization uses appropriate feedback mechanisms to ensure game quality. | | | | | | |
| 7 | The organization always develops a proper marketing plan and strategy to gain competitive advantage. | | | | | | |
| 8 | The organization is able to maximize market size and its growth over time. | | | | | | |
| **Innovation** | | | | | | | |
| 9 | The organization is able to use innovative ideas successfully for game development and game level repositioning. | | | | | | |
| 10 | The innovations in games are aligned with existing business goals. | | | | | | |
| 11 | Reactive and proactive innovation in the game development process is supported by management. | | | | | | |
| 12 | Past innovative measures taken by the organization have helped in improving the game development and management process. | | | | | | |
| 13 | The organization believes that R&D investment can yield positive results in the near future. | | | | | | |
| **Relationship management** | | | | | | | |
| 14 | The organization has well-established mechanisms for data extraction, manipulation, and production for customer profiling, profitability analysis, and retention modeling. | | | | | | |
| 15 | The organization participates in online gaming communities to identify player concerns. | | | | | | |
| 16 | The organization is able to retain players for long periods. | | | | | | |
| 17 | The organization has established a balanced player- and game-centered strategy for game development. | | | | | | |
| 18 | The organization is able to attract new players and retain existing ones using innovative targeted methods and personalized communication. | | | | | | |
| 19 | The organization is using a user integration strategy for game development. | | | | | | |
| **Time to market** | | | | | | | |
| 20 | Games are launched in the market before competitors' games. | | | | | | |
| 21 | The organization regularly studies and researches development updates, market reviews, and game publishing schedules to build awareness of market needs and trends. | | | | | | |
| 22 | The organization publishes games in response to competitors' actions. | | | | | | |
| 23 | Being first in the market helps to retain players and tends to attract new ones. | | | | | | |
| **Monetization strategy** | | | | | | | |

| 24 | The organization is able to achieve its financial objectives successfully. | | | | | |
|----|---------------------------------------------------------------------------|--|--|--|--|--|
| 25 | The organization is able to use cost-saving strategies successfully. | | | | | |
| 26 | The organization is able to acquire more players for less investment. | | | | | |
| 27 | The organization uses in-depth mechanics to maximize conversion rate and lifetime value in games. | | | | | |
| 28 | The organization can successfully build cross-platform offerings to reach players/consumers. | | | | | |
| **Brand name strategy** | | | | | | |
| 29 | The game development process of the organization is unique and different from its competitors in the market. | | | | | |
| 30 | New games and their latest versions are consistent with brand extensions. | | | | | |
| 31 | The latest game or its extended version attracts new customers and retains existing one because it is considered an improvement in a newer or existing game. | | | | | |
| 32 | The buying decision of the customer is based on brand name loyalty. | | | | | |
| 33 | Published games have one-to-one competition in the market. | | | | | |
| **Game business performance** | | | | | | |
| 1 | The organization was able to reduce the development time and cost of games over the last five years. | | | | | |
| 2 | The organization's sales have improved gradually over the last five years. | | | | | |
| 3 | The organization's financial analysis shows progressive growth over the last five years. | | | | | |
| 4 | Players' purchasing decisions are influenced by our brand-name game. | | | | | |
| 5 | The organization has been able to reduce significantly the number of competitors over the last five years. | | | | | |
| 6 | The organization is considered as a pioneer in the digital game industry rather than as a follower. | | | | | |
| 7 | Customer satisfaction and loyalty ratings have increased over the last five years. | | | | | |
| 8 | The business goals of the organization have been successfully accomplished. | | | | | |